\documentclass[%
 aip,
sd,
 amsmath,amssymb,
preprint,%
]{revtex4-1}
\usepackage{graphicx}
\usepackage{dcolumn}
\usepackage{bm}

\usepackage[format=plain, justification=raggedright, aboveskip=0cm, belowskip=0cm]{caption}
\usepackage[utf8]{inputenc}
\usepackage[T1]{fontenc}
\usepackage{mathptmx}
\begin{document}


\title{Time-resolved diffraction with an optimized short pulse laser plasma X-ray source}



\author{M.\ Afshari}
\author{P.\ Krumey}
\author{D.\ Menn}
\author{M.\ Nicoul}
\author{F.\ Brinks}
\author{A.\ Tarasevitch}
\author{K.\ Sokolowski-Tinten}\email[]{Klaus.Sokolowski-Tinten@uni-due.de}
\affiliation{Faculty of Physics and Center for Nanointegration Duisburg-Essen, University of Duisburg-Essen, Lotharstrasse 1, 47048 Duisburg, Germany}


\date{\today}

\begin{abstract}

We present a set-up for time-resolved X-ray diffraction based on a short pulse, laser-driven plasma X-ray source. The employed modular design provides high flexibility to adapt the set-up to the specific requirements (e.g. X-ray optics, sample environment) of particular applications. The configuration discussed here has been optimized towards high angular/momentum resolution and uses K$_{\alpha}$-radiation (4.51 keV) from a Ti wire-target in combination with a toroidally bent crystal for collection, monochromatization and focusing of the emitted radiation. $2\times 10^5$ Ti-K$_{\alpha1}$ photons per pulse with $10^{-4}$ relative bandwidth are delivered to the sample at 10 Hz repetition rate. This allows for high dynamic range ($10^4$) measurements of transient changes of the rocking curves of materials as for example induced by laser-triggered strain waves.
\end{abstract}


\maketitle
\section{\label{sec:Intro}Introduction}
By combining atomic scale spatial and temporal resolution ultrafast time-resolved diffraction using short X-ray or electron pulses provides direct access to the atomic motions in materials on their natural time-scale, i.e.\ femtoseconds to picoseconds. This relatively new field of {\it structural dynamics} has seen tremendous progress in recent years mainly driven by the development of new sources (e.g.\ refs.\ \citenum{elsaesser14,miller14,weathersby15,schoenlein19} and references therein). In the case of X-rays the current standard is set by X-ray free electron lasers, which exhibit extreme brightness, ultrashort pulse duration (currently down to the few fs level) and spatial coherence offering spectacular new opportunities\cite{bostedt16}. While more and more of these large-scale facilities are getting operational and available to users, access is highly competetive and still very limited. Therefore, as an alternative lab-scale approaches are still being pursued and developed. Among those the hard X-ray emission of short-pulse laser-produced plasmas has found wide-spread use for radiography/imaging (e.g.\ refs.\ \citenum{krol97a,toth07,chen10}), time-resolved X-ray absorption (e.g.\ refs.\ \citenum{raksi96,reich07,dorchies11,miaja16,anwar19}) and, in particular, for ultrafast diffraction (e.g.\ refs.\ \citenum{rosepetruck99,siders99,rousse01,sokolowski01,sokolowski03,bargheer04,korff07,quirin12,stingl12,juve13,pudell18}). In fact, the first time-resolved X-ray diffraction experiment with sub-picosecond resolution has been performed at such a source\cite{rischel97}.

By focusing a femtosecond laser pulse at intensities in excess of $10^{16}$ Wcm$^{-2}$ onto the surface of a solid target high temperature plasmas with near-solid-density can be generated \cite{Gibbon2005} which represent an efficient source of hard X-rays \cite{Kuhlke87, murnane91}. The emitted radiation contains continuum and characteristic line emission from the thin surface plasma layer as well as the "cold" solid behind. Due to collisionless interactions\cite{Brunel1987, Gibbon92, Teubner93, teubner_ph_p_96, teubner_j_phy_96, reich00, ziener2002, reich03, weisshaupt15} (resonance absorption and/or vacuum heating) between the created plasma and the laser pulse, a fraction of the plasma electrons is accelerated to kinetic energies of several tens of keV, much higher than the thermal energy of the rest of the plasma electrons (several hundreds of eV)\cite{rousse94}. These "hot" electrons generate Bremsstrahlung and characteristic line emission very similar to a conventional X-ray tube by penetrating into the cold solid underneath the plasma layer. Since these high energy, "hot" electrons are a result of the direct laser-plasma interaction, the X-ray pulse duration can be comparable to the driving laser pulse duration \cite{reich00, zamponi10}.

The efficiency of X-ray production critically depends on the hot electron distribution (their energy and number) and thus on the details of the laser-plasma interaction, which can be controlled through the laser parameters such as wavelength, intensity, angle of incidence and laser polarization as well as the properties of the created plasma (e.g.\ scale length).

For example, the K-shell ionization cross-section is maxiumum at electron energies of a few times the K-shell ionization energy of a given material.\cite{powell76} Consequently, optimum K$_{\alpha}$ production can be expected if the average energy of the "hot" electron distribution peaks in this range \cite{reich00, Ewald2002, eder00}. This average energy, often described by an effective "hot" electron temperature $T_h$, scales with the ponderomotive energy, i.e.\ $T_h \propto I_0\cdot\lambda^{2}$, where $I_0$ is the laser intensity and $\lambda$ its wavelength \cite{Meyerhofer93, Jiang95, Gibbon96_pl_fusion, Feurer2001}. Therefore, laser wavelength as well a laser intensity are control parameters to improve the efficiency of X-ray production\cite{weisshaupt14} or to push the X-ray emission to higher energies\cite{fourmaux16, azamoum18b, azamoum18}.

Similarly important is the plasma scale length reached at the peak of the laser pulse where the intensity is maximum \cite{Gibbon92}. Since plasma formation occurs already at intensities in the range of $10^{13}$ Wcm$^{-2}$, the conversion efficiency is very sensitive to the temporal structure of the rising edge of the laser pulse. If the laser-pulse contrast ratio (LPCR) is low (e.g. due to imperfect stretching and recompression of the laser pulses in the typically used chirped-pulse-amplification - CPA - laser systems or due to pre-pulses and/or amplified spontaneous emission - ASE) plasma formation and expansion occur well before the pulse maximum. In some cases \cite{ziener2002, reich03, Zhavoronkov2005, Khattak2006} "passive" optimization of X-ray production has been achieved through the inherent time structure of the given drive laser. In contrast, at laser systems with high LPCR controlled pre-pulse schemes have been employed to actively improve laser-driven plasma X-ray sources\cite{bastiani97, Schlegel99, eder00, Nakano2001, sokolowski02, Lu2009}.

Based on our previous detailed investigations of K$_{\alpha}$ X-ray production \cite{Lu2009}, we present here a set-up for time-resolved X-ray diffraction based on an optimized laser-driven plasma Ti-K$_{\alpha}$ X-ray source. We employ a modular design that provides high flexibility with respect to the specific requirements (e.g. X-ray optics, sample environment) of particular applications. The configuration discussed here has been optimized towards high angular/momentum resolution by using a toroidally bent crystal for collection, monochromatization and focusing of the emitted radiation.

The paper is organized as follows: In section \ref{sec:setup} we give a brief description of the technical features of the set-up. The main section discusses the spectral characterization of the X-ray source (\ref{sec:spectrum}), the optimization of its K$_{\alpha}$ yield  (\ref{sec:optimize}), the characterisation of the toroidally bent crystal used for focusing and monochromatization (\ref{sec:optics}), and static as well as time-resolved diffraction experiments on "test" samples to demonstrate its performance (\ref{sec:static}). Finally, section \ref{sec:summary} summarizes the properties of the setup.

\section{\label{sec:setup}Experimental realization}
The laser used as the driver for the X-ray source is a home-build CPA Ti:Sapphire laser system, including an oscillator, eight-pass pre-amplifier and a four-pass booster amplifier. The system provides pulses with $<$100 fs pulse duration at a center wavelength of 800 nm, and a pulse energy of approx.\ 150 mJ at 10 Hz repetition rate. The pulses exhibit a high LPCR of about $10^{7}$ at 2 ps ahead of the pulse peak; the LPCR to ASE is better than $10^{8}$.

First the incoming laser beam is split into a "main-pulse" and a weaker "pre-pulse" beam by using a mirror with a hole in the center ("holey-mirror") as beam splitter. While the "main-pulse" represents the actual X-ray driver, the "pre-pulse" is used to generate a pre-plasma to enhance X-ray generation (see section \ref{sec:optimize}). After introducing a suitable delay these two beams are recombined by a second "holey-mirror" beam splitter. Before, a third beam, which serves as the "pump pulse" to excite the sample under study (seperate delay control), is split from the "main-pulse" by an off-center "holey-mirror" beam splitter.

\begin{figure}[h]
\includegraphics[width=1\linewidth]{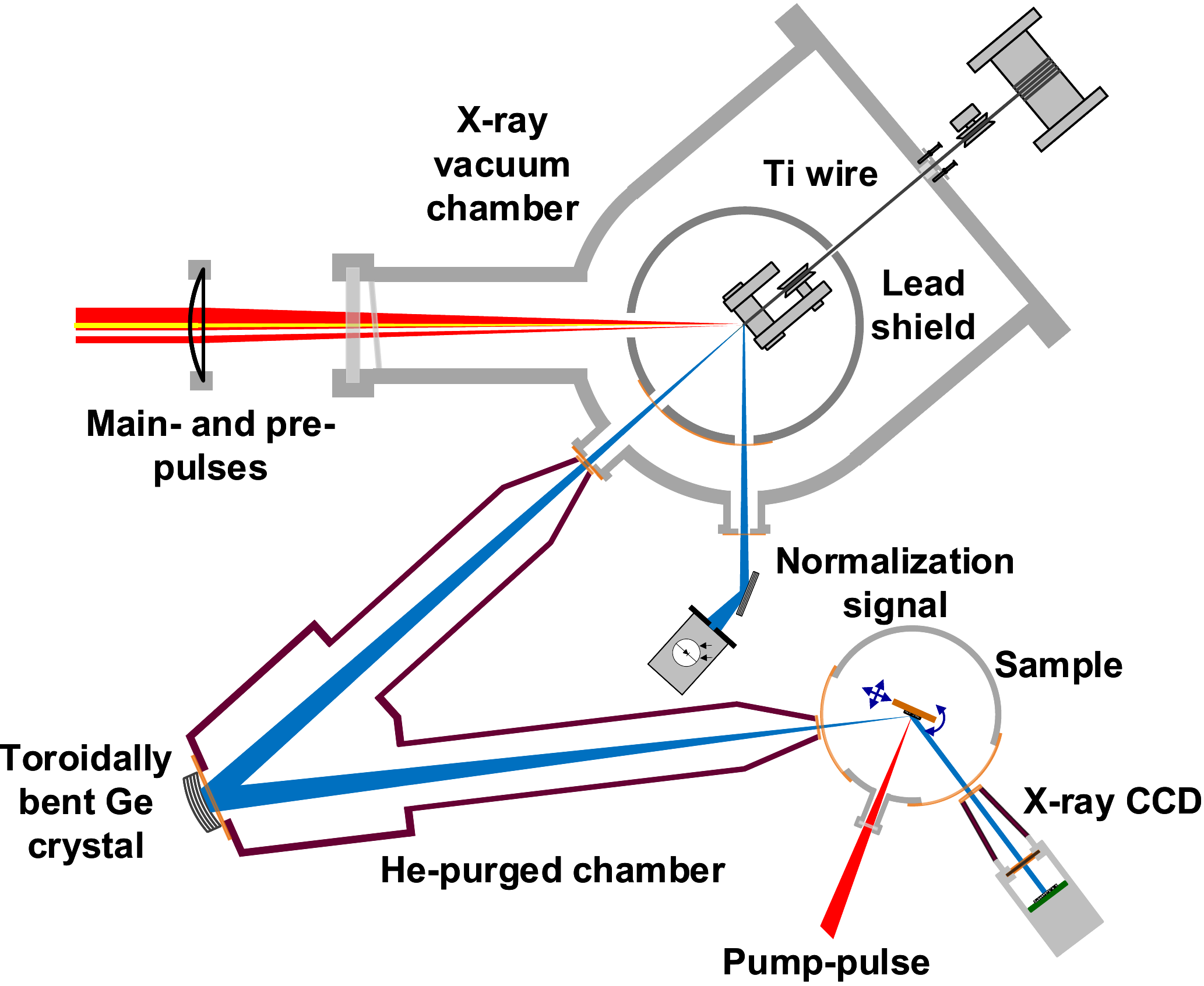}
\caption{Schematic of the optical-pump - X-ray-probe set-up. \label{fig1}}
\end{figure}

"Main-" and "pre-pulse" are guided collinearly to a small vacuum chamber (see Fig. \ref{fig1}; pressure $\approx$ 10$^{-2}$ mbar) and focused by a plano-concave lens with a focal length of 30 cm onto the surface of a Ti wire with a diameter of 250 $\mu$m. The beam diameters at the focus of the lens are 25 $\mu$m and 60 $\mu$m for "main-" and "pre-pulse", respectively. With the given pulse energy this results in maximum intensities on the wire of close to $10^{17}$ Wcm$^{-2}$ for the "main pulse" and $\approx 10^{15}$ Wcm$^{-2}$ for the "pre-pulse". Due to their high intensity both pulses induce material ablation in the irradiated area and a fresh target has to be provided for each laser pulse (pair). Therefore, the wire is continously pulled over high-precision, ball-bearing mounted guides using a motor with adjustable torque installed outside the vacuum chamber, resulting in a positional stability of about $\pm$ 5 $\mu$m in all directions.

For radiation safety purposes and in order to eliminate any hard X-ray background the wire-target assembly is enclosed by a lead-housing with minimized laser input and X-ray output openings. Additional lead shielding is attached to the inner walls of the stainless-steel vacuum chamber. Under normal operating conditions this results in a radiation level below 1 $\mu$Sv/h at 10 cm distance from the chamber.

As in a conventional X-ray tube the X-ray emission of the plasma occurs spatially incoherent into the full solid angle. Therefore, suitable X-ray optics are required to collect and refocus the radiation of the plasma onto the sample under study. Here we use a toroidally bent Ge crystal in a 1:1 imaging Rowland circle geometry\cite{missalla99,nicoul05}. The geometry is chosen such that the Bragg-condition is fulfilled over the whole area of the mirror resulting in very high reflectivity for a fraction of the bandwidth of the K$_{\alpha}$-emission. As discussed in more detail below a focus with a diameter of about 80 $\mu$m (FWHM), which contains up to 2$\times 10^5$ X-ray photons per pulse with a spectral bandwidth of approx.\ 0.43 eV centered on the K$_{\alpha1}$ line has been achieved at our source using this mirror.

With our modular scheme only the X-ray source needs to be in vacuum to avoid non-linearities in air due to the high intensity of the focussed laser beam. The other parts of the setup (X-ray optic, sample stage, detector) are separated from the X-ray source chamber and can be flexibly moved/exchanged if required. For example, to enable experiments at low temperature a cryostat with a small sample vacuum chamber can be inserted (as schematically depicted in Fig. \ref{fig1}). Alternatively we use an in-air sample manipulator/goniometer for room-temperature measurements, which allows for larger or multiple samples. Since the Ti-K$_{\alpha}$ radiation at 4.51 keV is significantly absorbed in air (1/e absorption length $\approx$16 cm \cite{henke93}), He-purged beam tubes are placed between the different components to minimize absorption.

A thinned, back-illuminated Si CCD (Princeton Instrument PI-MTE:1300B) is used as detector for the X-rays diffracted by the sample. This detector exhibits a quantum efficiency of 55$\%$ at 4.51 keV\cite{calib} and a chip area of 26.8 x26 mm$^2$ (1340 $\times$ 1300 pixels of 20 $\times$ 20 $\mu$m$^2$ size).

To account for the variation of the X-ray flux due to both long-term drifts and short-term fluctuations \cite{schick-norm12}, a "direct" normalization \cite{schick-norm12, zahng-norm14, hofer-norm16, holtz-norm2017} scheme has been implemented in which a GaAs crystal is properly placed at a second output of the X-ray source chamber and the integrated diffraction signal of its (111)-reflection is monitored by a large area (diameter 10 mm) X-ray sensitive avalanche photo-diode (APD). This allow to normalize the diffraction signals recorded by the CCD with an accuracy of better than 2$\%$.

The angle $\alpha$ between the optical pump and the X-ray probe beam in combination with the finite X-ray beam size lead to a variation of the relative arrival time at the sample surface limiting the temporal resolution of the experiment. For the current geometry ($\alpha \approx$ 50$^{\circ}$) this results in a temporal smearing of about 0.45 ps at a Bragg-angle of 20$^{\circ}$, which reduces to 0.2 ps for larger Bragg-angles. This is still sufficient for the investigation of transient strain effects - the application this set-up has been optimized for - which occur on {\it acoustic}, ps time-scales.

%
\section{\label{sec:source}Set-up characterization and -optimization}
This section discusses the detailed characterization and optimization of the setup. This includes measurements of the source spectrum, our efforts to maximize the K$_{\alpha}$ yield, as well as the performance characterization of the bent crystal mirror in terms of efficiency, focusing capability and bandwidth.

\subsection{\label{sec:spectrum}Spectral characterization}
\begin{figure}[h]
\includegraphics[width=1\linewidth]{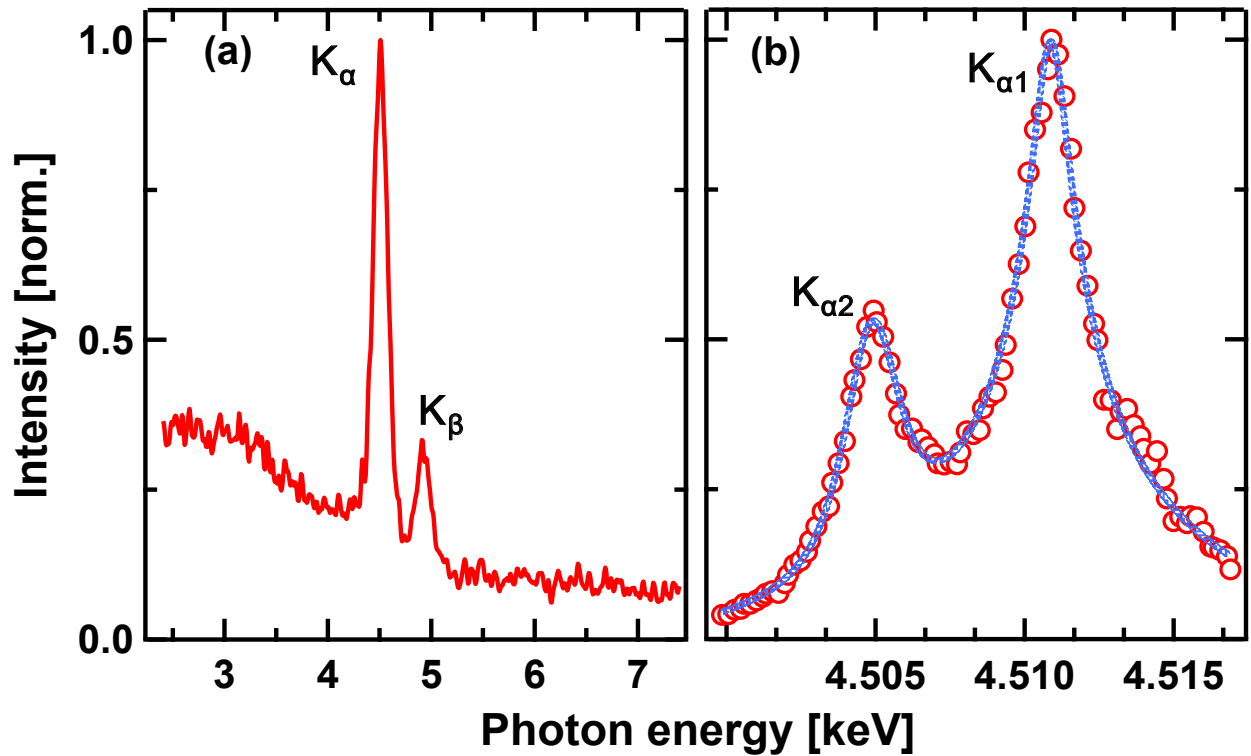}
\caption{Spectra of the plasma X-ray emission: (a) Low resolution spectrum measured by the X-ray CCD operating in photon counting mode. (b) High resolution spectrum of Ti-K$_{\alpha}$ emission measured by rocking the toroidally bent mirror (red circles; the blue line represents a guide to the eye).\label{fig2}}
\end{figure}

Spectral characterization has been done in two steps. First, the Si CCD was placed between the source and the bent crystal mirror and operated in photon-counting mode by drastically reducing the X-ray flux through a reduction of the drive laser power. In this mode the detector acts as a spectrometer since a single X-ray photon produces a charge in the detecting pixel that is proportional to the photon energy\cite{zamponi05}. Thus a histogram of the signals in all the pixels of the CCD represents the spectrum of the detected radiation. A typical result is shown in Fig.\ \ref{fig2}(a).

The measured spectrum is characterized by a broad continuum and two line emission features, which represent the K$_{\alpha}$ and K$_{\beta}$ emission of Ti. It needs to be stressed that the apparent continuum at energies below the K-shell emission lines is only partly due to Bremsstrahlung since at these photon energies there is a non-negligible probability that the charge generated by a K$_{\alpha}$ or K$_{\beta}$ photon is shared between two or more pixels\cite{zamponi05}. Due to the limited energy resolution of about 150 eV, the CCD is not able to resolve spectral fine structures, namely the spin-orbit split K$_{\alpha1}$ and K$_{\alpha2}$ lines (energy seperation 5.98 eV). Therefore, in a second step the toroidally bent Ge crystal, which provides a spectral resolution of $\approx$0.43 eV (see section \ref{sec:optics}), was utilized as a scanning spectrometer (by changing the incidence angle) to precisely measure the emitted K$_{\alpha}$ spectrum. As depicted in Figure \ref{fig2}(b), the K$_{\alpha1}$ and K$_{\alpha2}$ lines are completely resolved. Their width was measured as $\approx$3.1 eV, which is broader than the reported natural Ti-K$_{\alpha}$ linewidth of about 1.45 eV and 2.13 eV\cite{Salem76}, respectively. This broadening has been observed before\cite{Sengebusch2009, arora2011} and attributed to emission contributions from atoms in higher ionization states.

\subsection{\label{sec:optimize}Source optimization}
As discussed before, the drive laser intensity as well as the plasma scale length have a major influence on X-ray generation and thus the K$_{\alpha}$-flux available in a diffraction experiment.

\begin{figure}[h]
\includegraphics[width=1\linewidth]{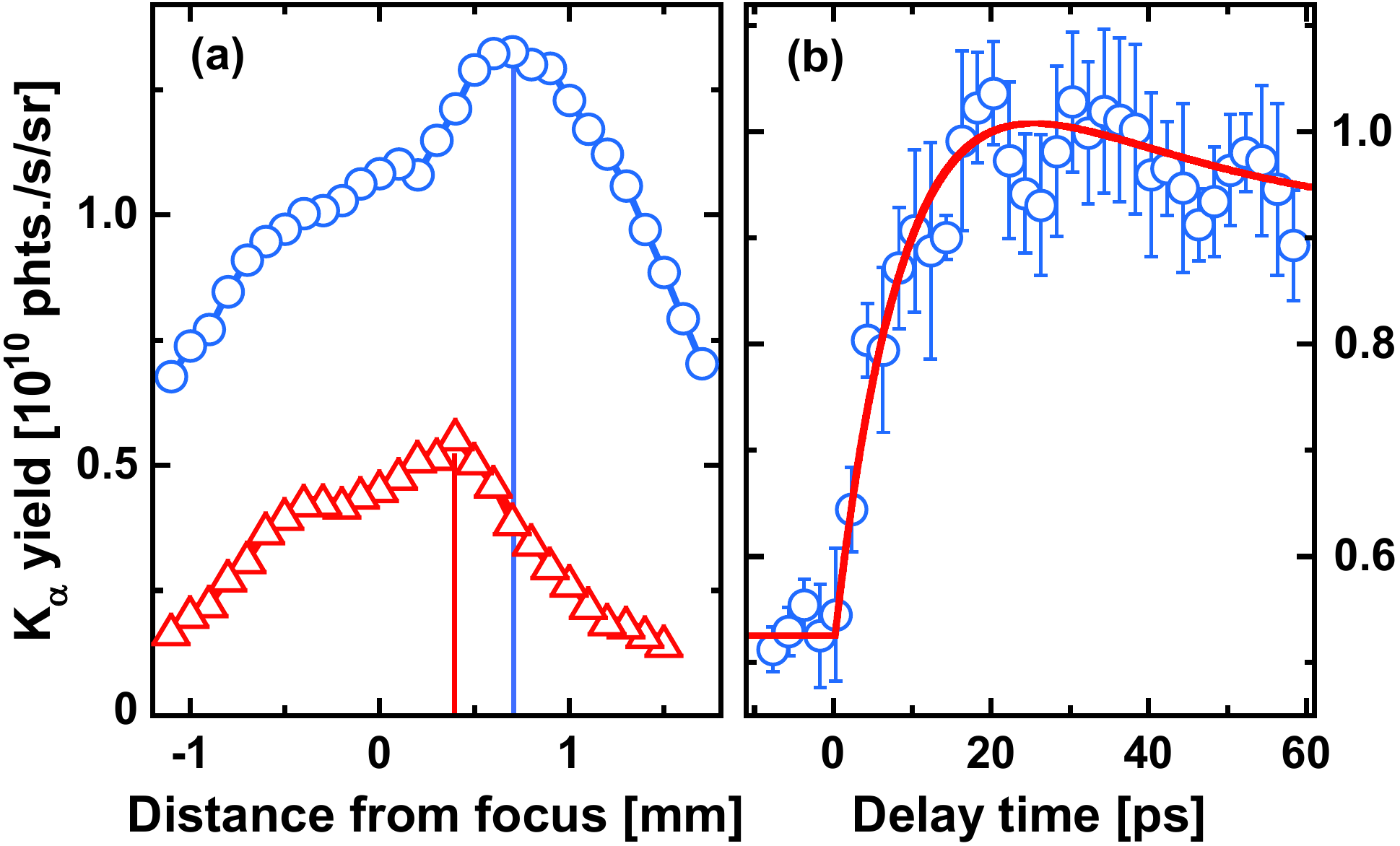}
\caption{Optimization of the K$_{\alpha}$ X-ray production. (a) K$_{\alpha}$ yield as a function of the relative position of the laser focus with respect to the surface of the Ti wire, without (red triangles) and with (blue circles) "pre-pulse" (at optimum delay, see (b)). (b) K$_{\alpha}$ yield as a function of delay time between "pre-" and "main-pulse" (blue circles; the red line is a guide to the eye.)}
\label{fig3}
\end{figure}

A straight-forward way to vary the intensity without changing the laser energy is to change the distance between the focusing lens and the wire, thus changing the laser spot size on the wire by moving it in and out of the focus, where the intensity is highest.  The red data points in Fig.\ \ref{fig3}(a) show the total K$_{\alpha}$ yield \cite{yield} as a function of the relative lens position (zero marks the position with the wire in the focus) for the case, when only the "main-pulse" is used for X-ray generation. It can be clearly seen that the maximum K$_{\alpha}$ yield is not obtained for the highest available intensity, but with the wire approx.\ 0.4 mm before the focus and thus at an intensity below $10^{17}$ Wcm$^{-2}$. This is in agreement with previous observations\cite{eder00, fill02, sokolowski02, reich03, Lu2009} and also with results of theoretical calculations\cite{reich00} which predict a maximum K$_{\alpha}$ yield for Ti at intensities of a few times $10^{16}$ Wcm$^{-2}$.

Due to the high LPCR of the laser system used here, only a short scale length pre-plasma is created by the leading edge of a single pulse. Therefore, the collisionless coupling of laser energy to the plasma and consequently the X-ray production is not optimum \cite{Gibbon92}. We have shown previously\cite{Lu2009} that it is possible to maximize the K$_{\alpha}$ flux by creating a pre-plasma with the angle of incidence-dependent optimum plasma scale length using a controlled "pre-pulse" with a suitable negative delay with respect to the X-ray generating "main-pulse". We apply this approach here. The "pre-pulse" had a maximum intensity of almost $10^{15}$ Wcm$^{-2}$ with the wire in the focus of the lens. The measured K$_{\alpha}$ yield versus delay time is depicted in Fig.\ \ref{fig3}(b). Positive delay times mean that the pre-pulse arrives earlier than the main-pulse, so the plasma scale length increases with delay time. As expected, the K$_{\alpha}$ flux is enhanced at positive delays, when the plasma generated by the pre-pulse has expanded. A maximum yield increase of about a factor of two is reached at approximately 20 ps. In line with our previous results\cite{Lu2009} a relatively long scale length pre-plasma and thus a long delay between pre- and main-pulse is required due to the near-normal incidence of the laser on the wire, which has been chosen to minimize the effects of fluctuations of the wire position as well as the laser pointing.

The data shown in Fig.\ \ref{fig3}(b) have been measured at the lens position which resulted in maximum K$_{\alpha}$ production {\it without} the "pre-pulse".\cite{prepulse} As illustrated by the blue data points in Fig.\ \ref{fig3}(a) we were able to improve the X-ray yield with "pre-pulse" further by reducing the intensity of both, the "pre-" and the "main-pulse", through an increase of the wire-focus distance. At optimum conditions a maximum Ti K$_{\alpha}$ flux of more than $1.3\times 10^{10}$ photons s$^{-1}$ sr$^{-1}$ was achieved, which corresponds to $\approx 2\times 10^5$ photons per pulse delivered to the sample.

\subsection{\label{sec:optics}X-Ray optics}
Almost at any X-ray source X-ray optical elements are used to direct, focus or more generally manipulate the radiation for an intended application. For example at accelerator based X-ray sources like synchrotrons and X-ray free electron lasers, which usually exhibit well collimated and often highly monochromatic beams, lenses \cite{lengeler05} as well as curved mirrors based on total reflection at grazing incidence \cite{yumoto12} are employed. In contrast, the spatially incoherent 4$\pi$-emission of laser-plasma based X-ray sources require optics, which allow to collect the radiation over a sufficiently large solid angle and deliver it to the sample in a suitably shaped beam (focused, collimated). Bent crystals, multi-layer mirrors and capillary optics have been utilized for this purpose\cite{missalla99,nicoul05,bargheer05,shymanovich08,rathore17,schollmeier18}.

Among those, bent crystals provide the highest spectral purity\cite{bargheer05}. Since the spectral bandwidth determines the angular/momentum resolution of a diffraction experiment, such an optic has been chosen for the current set-up. They are based on Bragg diffraction and can achieve a high reflectivity over a large area, since the lattice planes are parallel to the geometrical surface of the either spherically or toroidally bent crystals\cite{missalla99}.

\begin{figure}[h]
\includegraphics[width=1\linewidth]{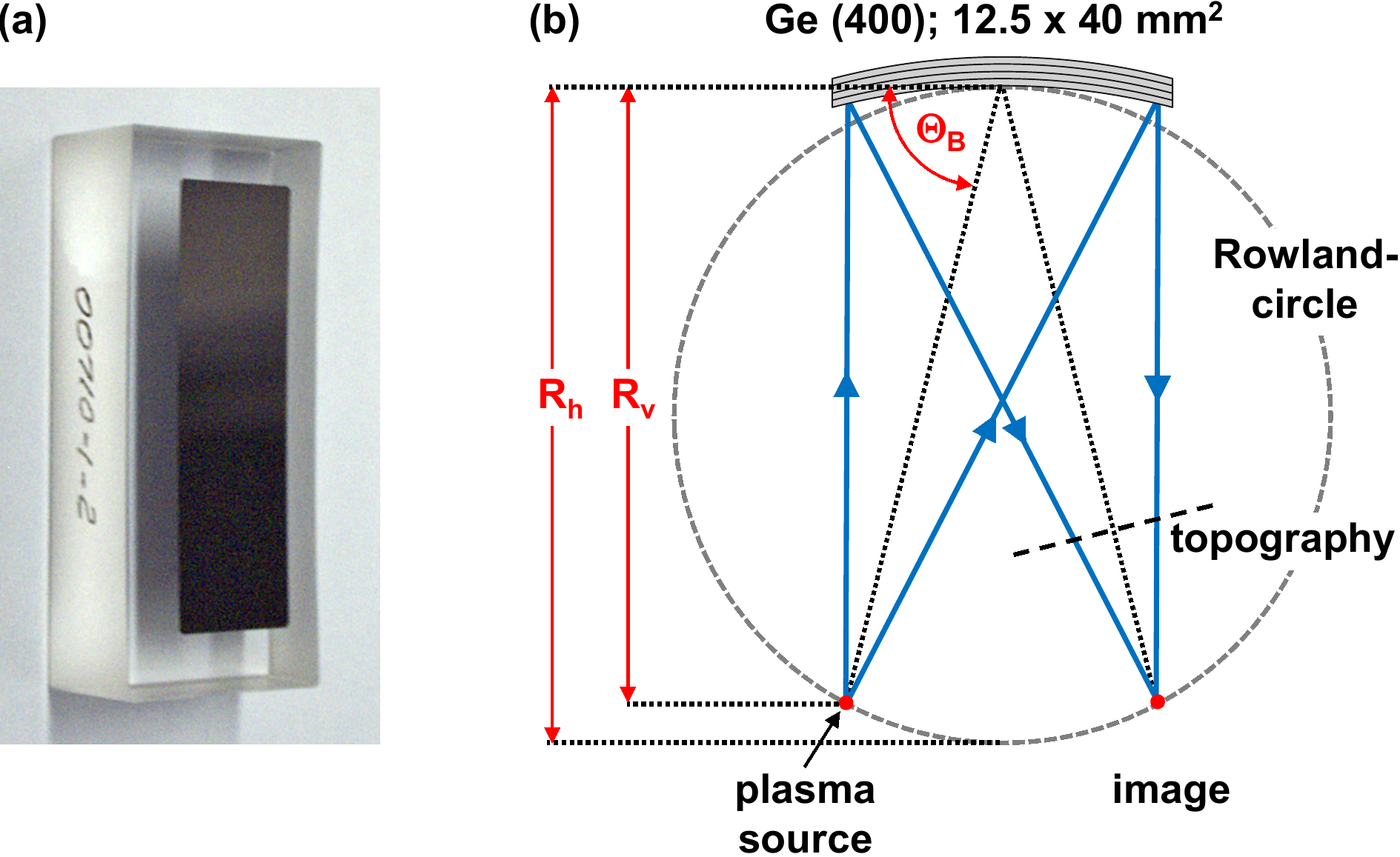}
\caption{(a) Photograph of the torodially bent crystal X-ray mirror. (b) Schematic of the Rowland circle geometry resulting in a 1:1 imaging of the plasma source. The dashed line labelled "topography" marks the position of the X-ray CCD used to measure the reflectivity profile over the mirror surface (Fig.\ \ref{fig5}(a)).\label{fig4}}
\end{figure}

In the current setup we employ the (400)-reflection of a toroidally bent Ge-crystal with (100)-orientation. It has been fabricated by INRAD inc.\cite{inrad_x} to our specifications and technical details have been discussed by Nicoul et al.\cite{nicoul05}. In brief, a 12.5 mm wide, 40 mm high, and 90 $\mu$m thick Ge crystal is bound to a toroidally shaped glass substrate (see photograph in Fig.\ \ref{fig4}(a)). Such a toroidally bent crystal mirror provides a quasi-monochromatic 1:1 image of a point-like source if source and image are located on the so-called Rowland-circle, as depicted in Fig.\ \ref{fig4}(b), such that the vertical and horizontal bending radii $R_V$ and $R_H$, respectively, satisfy the condition $\frac{R_V}{R_H} = \sin^{2}\theta_B$, with $\theta_B$ being the Bragg angle for the required X-ray wavelength (here $\theta_B = 76.32^o$ for Ti K$_{\alpha1}$).

\begin{figure}[h]
\includegraphics[width=1\linewidth]{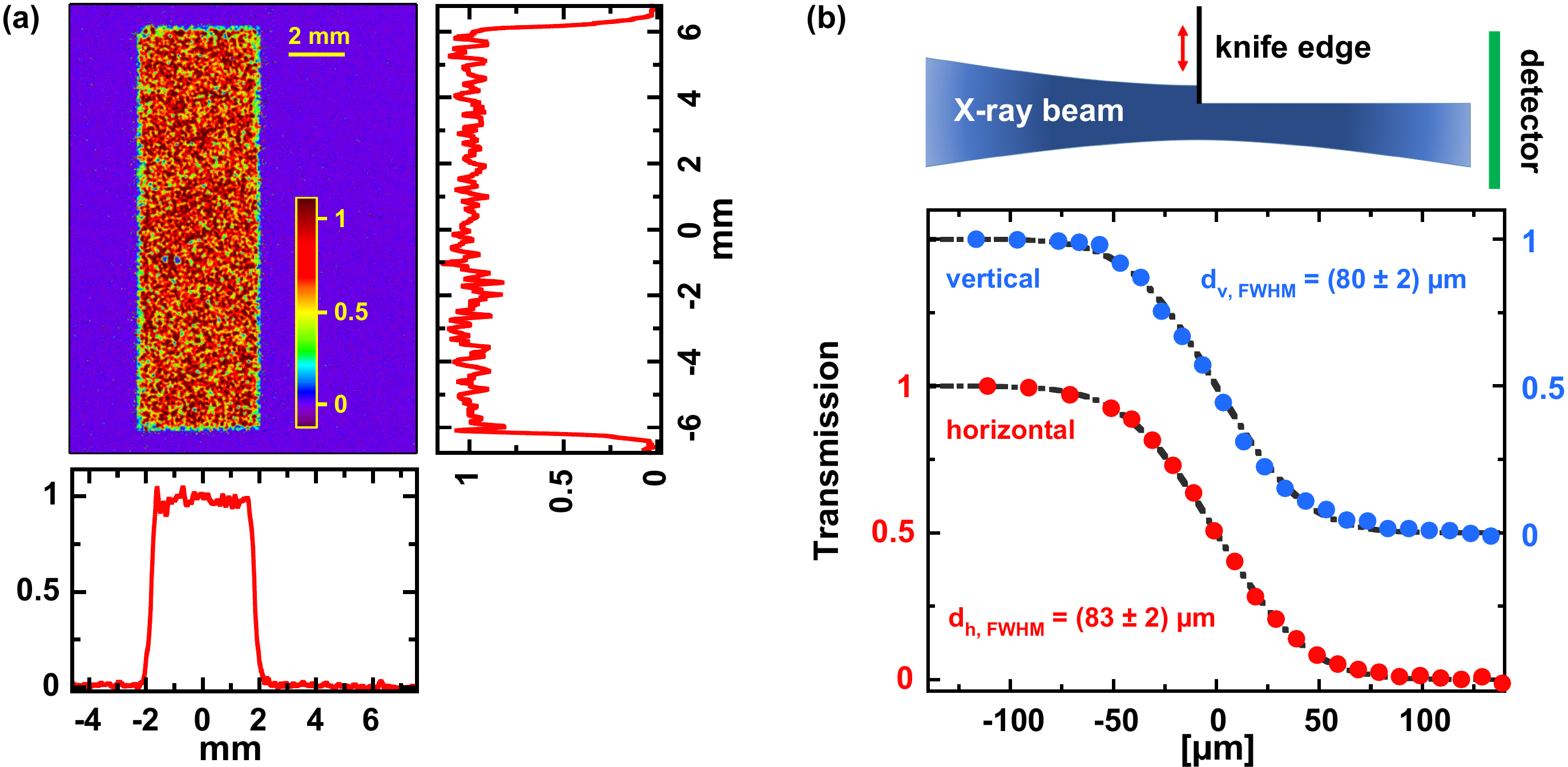}
\caption{(a) X-ray "topography" (spatial reflectivity distribution) of the bent crystal: False color image and vertical (right) and horizontal (bottom) cross sections. (b) Results of knife edge scans at the best focus position: Normalized K$_{\alpha1}$ transmission as a function of vertical (blue) and horizontal (red) blade position, respectively. The black dash-dotted curves represent error-function fits corresponding to the Gaussian beam spot sizes (FWHM) noted in the graph.}
\label{fig5}
\end{figure}

To ensure both, highest mononchromaticity and a homogeneous reflectivity across the mirror surface the source needs to be accurately positioned on the Rowland-cicle. Therefore, the source-mirror distance as well as the Bragg-angle (see also Fig.\ \ref{fig2}(b)) have been carefully adjusted by monitoring the intensity distribution of the reflected/diffracted K$_{\alpha}$ radiation with the X-ray CCD placed between mirror and sample/image position (marked by the dashed line labelled as "topography" in Fig.\ \ref{fig4}(b)). Figure \ref{fig5}(a) shows the reflectivity distribution and its vertically and horizontally averaged cross-sections for optimum adjustment. Despite fluctuations resulting from the short detector integration time (corresponding to a relatively low average number of photons per pixel), these data evidence a homogeneous reflectivity across the entire mirror surface. In the image plane ("focus" of the X-ray mirror) this transfers into a monochromatic and homogeneous intensity distribution as a function of angle over the full convergence range of 1.4$^{\circ}$ in horizontal direction (the dispersion direction of our set-up) and 4.5$^{\circ}$ vertically. In this configuration the complete angular dependence of the diffraction signal of a sample, i.\ e.\ its rocking curve, can be obtained without actually {\it rocking} (rotating) the sample.

Another critical point for optical pump - X-ray probe experiments concerns the exact determination of the focus/image position of such a mirror, because severe distortions of the angular distribution of the X-rays diffracted off the laser-excited area can occur if the sample under study is not properly positioned in the focus.\cite{shymanovich07} To precisely localize the focus, knife-edge scans using blades mounted on the sample stage exactly in the plane of the sample surface have been performed for different distances between mirror and sample (see top schematic in Fig. \ref{fig5}(b)). Results of such knife-edge scans for the best focus are depicted in the bottom part of Fig.\ \ref{fig5}(b), which shows the normalized "transmitted" signal as a function of the position of the horizontal (red) and vertical (blue) blade, respectively. The measured data can be described very well by an error function $T(x) = \frac{1}{2}\left(1 - erf\left(\frac{x}{x_0}\right)\right)$ (black dashed-dotted curves), where $x_0$ corresponds to the $1/e$-radius of a Gaussian beam. From these fits we determine the focus/image diameter (FWHM) to 83 $\pm$ 2 $\mu$m and 80 $\pm$ 2 $\mu$m in horizontal and vertical direction, respectively. This size represents the convolution between the imaging properties of the bent mirror and the X-ray source size (which we have not measured here) and is small enough to allow for a sufficient pump-probe spot size ratio.

\subsection{\label{sec:static}Static and Dynamic Diffraction}
In this section we present the results of static (without laser pumping) and time-resolved (with laser pumping) diffraction measurements to discuss treatment of the diffraction data, to characterize the angular/momentum resolution of the experiment, and to demonstrate the overall performance and sensitivity of the setup. For this purpose two different samples have been investigated, namley a (100)-oriented bulk GaAs crystal and a 180 nm thick, (111)-oriented Ge film, hetero-epitaxially grown on a (111)-oriented bulk Si substrate.\cite{hoegen94}

\begin{figure}[h]
\includegraphics[width=1\linewidth]{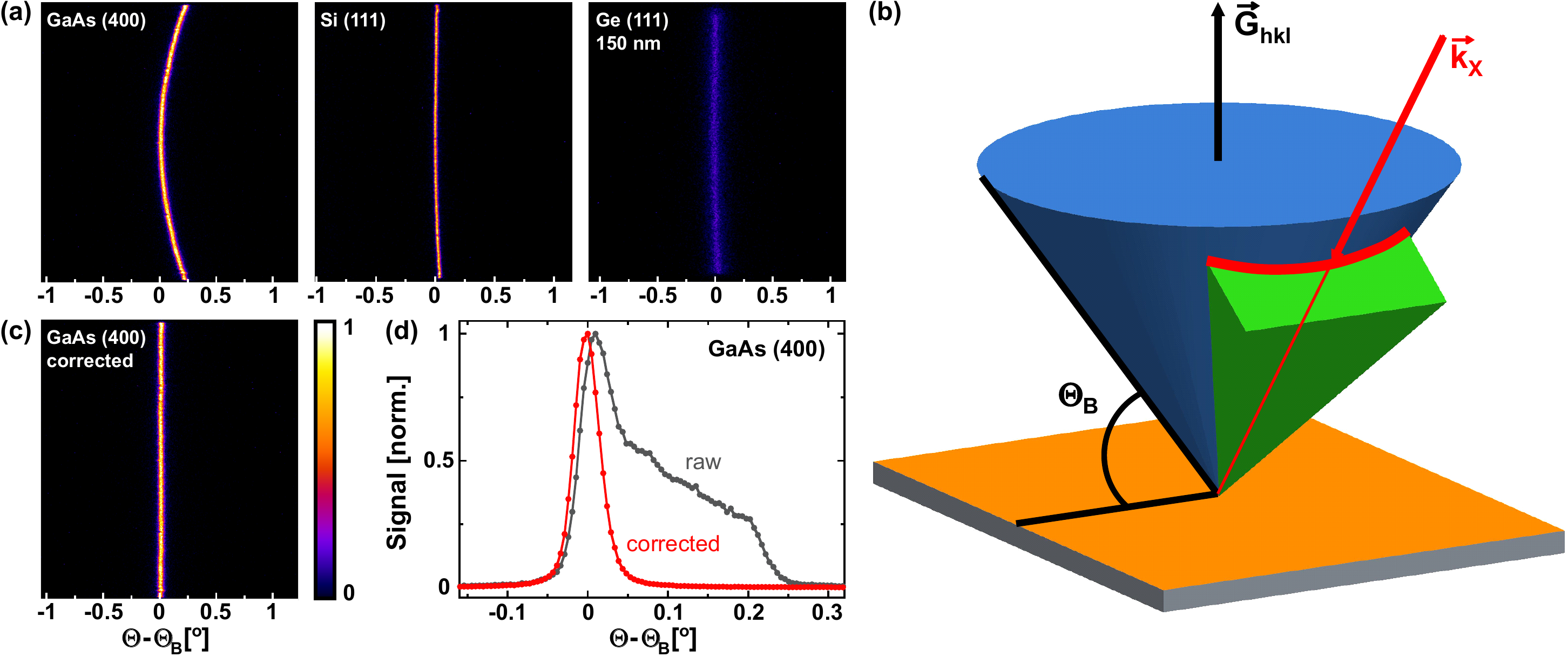}
\caption{(a): Raw detector images of (i) the (400)-reflection of a (100)-oriented GaAs-crystal, (ii) the (111)-reflection of a (111)-oriented Si-crystal, and (iii) the (111)-reflection of a 180 nm thick, (111)-oriented Ge film grown on the Si crystal. (b) Schematic of diffraction geometry. Blue: Kossel-cone corresponding to Bragg-reflection $\vec{G}_{hkl}$. Green: Cone of X-rays directed to the bent crystal mirror onto the sample with the center ray $\vec{k}_X$ adjusted that it fulfills the Bragg-condition and lies, therefore, on the Kossel-cone. Red curve: Line of intersection between the Kossel-cone and the cone of incident X-rays. (c) Diffraction image of the GaAs (400)-reflection after bending correction. (d) Angular dependence of the diffraction signal obtained by vertical integration of the diffraction images without (dark grey) and with (red) bending correction.\label{fig6}}
\end{figure}

Fig.\ \ref{fig6}(a) (top panel) shows raw detector images of (i) the (400)-reflection of the GaAs crystal, (ii) the (111)-reflection of the Si substrate, and (iii) the (111)-reflection of the 180 nm thick Ge film on-top the Si crystal, all obtained with an integration time of 1 minute (600 X-ray pulses) and without optical pumping. With an incident X-ray flux of about $2\times10^5$ K$_{\alpha}$ photons per pulse the detected integrated diffraction signal in photons per pulse is 240, 90, and 60 for the GaAs (400), Si (111), and Ge (111) reflections, respectively.

All diffraction patterns exhibit the shape of curved lines (most pronounced for the GaAs (400)). This is caused by the fact that a X-ray "beam" with a large convergence (1.4$^{\circ}$ horizonatlly and 4.5$^{\circ}$ vertically) is used. As depicted schematically in Fig.\ \ref{fig6}(b) all possible incident and diffracted X-rays for a particular reflection (hkl) lie on the so-called Kossel-cone (blue), which has a full opening angle of $180^{\circ} - 2\theta _B$ and an axis along the reciprocal lattice vector $\vec{G}_{hkl}$. From the Kossel-cone the X-rays (with center ray $\vec{k}_X$) focused by the bent mirror onto the sample surface (marked green in Fig.\ \ref{fig6}(b)) cut out a curved, line-shaped segment (red). Since the opening angle of the Kossel-cone decreases with increasing Bragg-angle the curvature of the diffraction pattern is strongest for the (400)-reflection of GaAs (Bragg-angle of 76.4$^\circ$).

In these images the horizontal axis corresponds to the "dispersive" direction and rocking curves are, in principle, obtained as horizontal cross sections after vertical integration of the diffraction pattern. However, the bending of the diffraction pattern lead to  distortions of the rocking curves. Therefore, to achieve the highest possible angular resolution, we applied a bending correction by fitting the curved diffraction line by a parabola which is then used to "unbend" the whole pattern, e.g.\ Fig.\ \ref{fig6}(c). The effect as well as the necessity for this bending correction is illustrated by Fig.\ \ref{fig6}(d), which shows the rocking curve of the GaAs (400) reflection without (grey curve) and with (red curve) bending correction. The rocking curve obtained from the uncorrected diffraction pattern is broadened and strongly asymmetrically deformed compared to the corrected case.

\begin{figure}[h]
\includegraphics[width=1\linewidth]{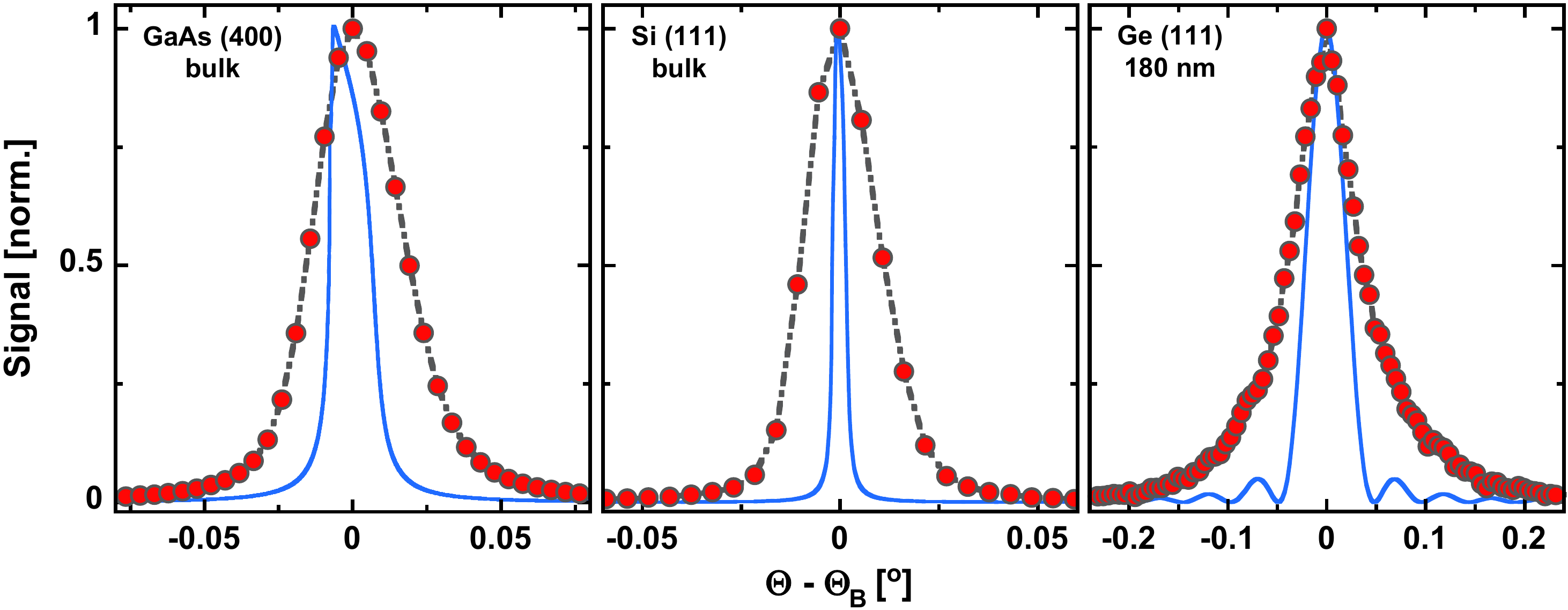}
\caption{Rocking curves of the GaAs (400)-reflection (bulk crystal, left, same data as in Fig.\ \ref{fig6}(d)), the Si (111)-reflection (bulk, middle), and the Ge (111)-reflection (180 nm film, right). Red-grey: Experimental data after bending correction; blue: Calculated rocking curves using the XCrystal-routine from the XOP-package\cite{XOP} (ver. 2.3)\label{fig7}}
\end{figure}

Fig.\ \ref{fig7} shows (red-gray) the rocking curves derived from the measured diffraction pattern in Fig.\ \ref{fig6} after bending correction. The experimental curves are compared to calculated rocking curves (blue) using the the XCrystal-routine from the XOP-package\cite{XOP} (ver. 2.3).

It is obvious that all experimental curves exhibit a larger width than the curves calculated for perfect crystals and a strictly monochromatic and fully collimated X-ray beam. The experimental rocking curve width $\Delta\Theta _{exp}$ has three different contributions, namely due to the finite X-ray spot size (converted into angle) on the sample $\Delta\Theta _{spot}$, due to the bandwidth of the radiation reflected by the bent mirror $\Delta\Theta _{bw}$, and due to the "natural" rocking curve width $\Delta\Theta _{rc}$ of the corresponding reflection (polarization averaged). We assume here $\Delta\Theta^2 _{exp} = \Delta\Theta^2 _{spot} + \Delta\Theta^2 _{bw}+ \Delta\Theta^2 _{rc}$.

Commercial wafers of GaAs and Si, as have been used here, exhibit almost perfect crystalline structure. Therefore their "natural" rocking curve should be close to the calculated ones. For the Si (111)-reflection we measure an angular width of $\Delta\Theta^{Si} _{tot} = $ 0.022$^\circ$ corresponding to 4 pixels or 80 $\mu$m on the detector and thus equal to the measured X-ray spot size on the sample (compare Fig.\ \ref{fig5}(b)). In this case the contributions from the natural rocking curve width $\Delta\Theta _{rc} = $ 0.003$^\circ$ and the bandwidth $\Delta\Theta _{bw} = $ 0.003$^\circ$ (see below) are negligible. In contrast, for the case of the thin Ge-film $\Delta\Theta _{rc}$ dominates the overall width. However, the measured width of 0.074$^\circ$ is significantly larger than the width of the calculated rocking curve (0.04$^\circ$). Moreover, the experimental rocking curve lacks the thickness fringes of the calculated curve. We attribute both observations to a finite mosaic spread in the hetero-epitaxially grown film. Finally, for the GaAs (400)-reflection, all three effects contribute similarly, which allows to determine the spectral bandwidth of the radiation reflected by the mirror. With a total width of $\Delta\Theta _{tot} =$ 0.034$^\circ$, a spot size contribution of $\Delta\Theta _{spot} =$ 0.019$^\circ$, and a "natural" rocking curve width of $\Delta\Theta _{rc} =$ 0.015$^\circ$, a bandwidth contribution of $\Delta\Theta _{bw} =$ 0.023$^\circ$ is obtained. This results in an energy bandwidth of $\Delta E_{mi} = \frac{\Delta\theta_{bw}}{\tan\theta_B}\:E_X \approx 0.43$ eV ($\theta_B$ = 76.5$^\circ$ GaAs (400) Bragg angle, $E_X =$ 4.51 keV X-ray photon energy) or a relative bandwith of $\approx$10$^{-4}$. This bandwidth is comparable, but larger than the "natural" bandwidth of the Ge (400)-reflection of a plane crystal of $\Delta E_{rc} =$ 0.27 eV. We attribute this to slight strain effects in the bent crystal, which will increase its bandwidth and decrease the peak reflectivity.\cite{uschmann93}

With the mirror bandwidth $\Delta E_{mi} = $ 0.43 eV, its acceptance solid angle $\Delta\Omega_{mi} = 1.92\times 10^{-3}$ sr, assuming a polarization-averaged peak reflectivity $R_{av} = 0.85$ (chosen somewhat smaller than the value of 0.92 for the plane crystal), and the measured Ti-K$_{\alpha1}$ linewidth of $\Delta E_{K_{\alpha1}} = $ 3.1 eV we can derive the mirror efficiency as $\eta = R_{av}\cdot\left(\Delta E_{mi}/\Delta E_{K_{\alpha1}}\right)\cdot\left(\Delta\Omega_{mi}/4\pi\right) \approx 1.8\times 10^{-5}$ (this value has been used to estimate the total $K_{\alpha}$-yield of our plasma X-ray source; see sec.\ \ref{sec:optimize}).

We finally present here exemplary time-resolved data obtained on the Ge/Si heterostructure after optical excitation of the Ge top layer with 100 fs, 800 nm laser pulses. Fig.\ \ref{fig8} shows the measured transient rocking curves (red) of the (111)-reflection of the 180 nm Ge overlayer (left column, linear scale) and the (111)-reflection of the bulk Si-substrate (right column, logarithmic scale) for three different pump-probe time delays. The grey curves represent the corresponding rocking curves of the unexcited sample measured at a pump-probe time delay of -15 ps, i.e. before arrival of the optical pump.

For the Ge film we observe a shift and broadening of the whole rocking curve towards smaller diffraction angles, indicating (an initially inhomogeneous) expansion of the lattice. In contrast, the main peak of the Si-rocking curve remains essentially unchanged, but develops shoulders, initially only on the high angle side (indicating compression), but later also on the low-angle side (indicating expansion). This behavior can be explained by strain waves\cite{thomson86}, which are triggered by the almost instantaneous increase of stress/pressure in the Ge-film upon its electronic excitation as well as the subsequent lattice heating.

Initially rarefaction waves are launched at both boundaries of the Ge film (the free surface as well as the Ge-Si interface), which propagate back and forth in the Ge-film while being partially transmitted into the Si-substrate on each round-trip. Expansion of the Ge-film is evidenced by the overall shift (and broadening) of the rocking curves towards smaller diffraction angles. This expansion leads to a compression of the Si substrate (shoulder of the Si rocking curve on the high angle side for 15 ps and 33 ps). At later times a train of bi-polar strain pulses\cite{thomson86, rosepetruck99, shymanovich09} develops resulting in shoulders/sattellites on both sides of the main Si (111)-peak.

The detailed strain evolution is determined by the complex interplay of electronic and thermal stress contributions,\cite{thomson86} which, as our measurements reveal, exhibit pronounced temporal and fluence dependencies. This we attribute to the dependence of the effective deformation potential, which determines the magnitude of the electronic stress, on the fluence- and time-dependent density of the laser-excited electron-hole plasma. While a detailed discussion of these processes is beyond the scope of this paper and will be presented in a separate publication\cite{afshari19}, we would like to stress that the high dynamic range (best visible for the logarithmically presented Si-data) of almost 10$^4$, enabled us to monitor even subtle changes of the rocking curves with high sensitivity and was key to separate and quantify the different stress contributions.

\section{\label{sec:summary}Discussion and Summary}
In summary, we have presented here a modular set-up for time-resolved "optical-pump - X-ray-probe" diffraction experiments which is based on a low repetition rate (10 Hz), laser-driven plasma K$_{\alpha}$ X-ray source. X-ray production with Ti as target material has been optimized by carefully adjusting the laser intensity as well as by employing a "pre-pulse"-scheme resulting in a total Ti K$_{\alpha}$ (4.51 keV) flux of up to $1.7\times 10^{11}$ photons per second into the full solid angle. By using a toroidally bent Ge (100) crystal to collect and refocus the K$_{\alpha}$ emission of the plasma, narrow bandwidth (0.43 eV; $10^{-4}$ relative) radiation with $\approx 2\times 10^{6}$ photons per second and a small spot size of $\approx$ 80 $\mu$m (FWHM) can be delivered to the sample. Table \ref{tab:setup_proper} summarizes the characteristics of the set-up.\cite{duration}

\begin{table}[htb]
\caption{\label{tab:setup_proper}Summary of the set-up properties.}
\begin{ruledtabular}
\begin{tabular}{lc}
 Parameter&Value\\
\hline
Laser wavelength& 800 nm\\
Laser pulse duration& 100 fs\\
Repetition rate& 10 Hz\\
"Main-pulse" intensity (max.) / diameter (focus)& $\sim$$10^{17}$ Wcm$^{-2}$ / $\sim$25 $\mu$m (FWHM)\\
"Pre-pulse" intensity (max.)/ diameter (focus)& $\sim$$10^{15}$ Wcm$^{-2}$ / $\sim$60 $\mu$m (FWHM)\\
Optimum delay between "main-" and "pre-pulse"& $\sim$20 ps\\
Ti-K$_{\alpha}$ yield& $\sim$1.3$\times 10^{10}$ photons s$^{-1}$ sr$^{-1}$\\
Bent mirror spectral bandwidth / relative bandwidth\footnotemark[1] & $\sim$0.43 eV / $\lesssim$10$^{-4}$\\
Bent mirror efficiency& $\sim$1.8$\times 10^{-5}$\\
X-ray convergence angle (hor. / vert.)& 1.4$^{\circ}$ / 4.5$^{\circ}$\\
X-ray focal spot diameter& $\sim$80 $\mu$m (FWHM)\\
Average X-ray spectral brightness\cite{Mills2005} (in the focus)& $\sim$2$\times 10^{6}$\\
&photons s$^{-1}$ mm$^{-2}$ mrad$^{-2}$ (0.1$\%$ bandwidth)$^{-1}$\\
\end{tabular}
\end{ruledtabular}
\footnotetext[1]{centered at Ti-K$_{\alpha1}$ $\equiv$ 4.51 keV.}
\end{table}

\begin{figure}[h]
\includegraphics[width=0.9\linewidth]{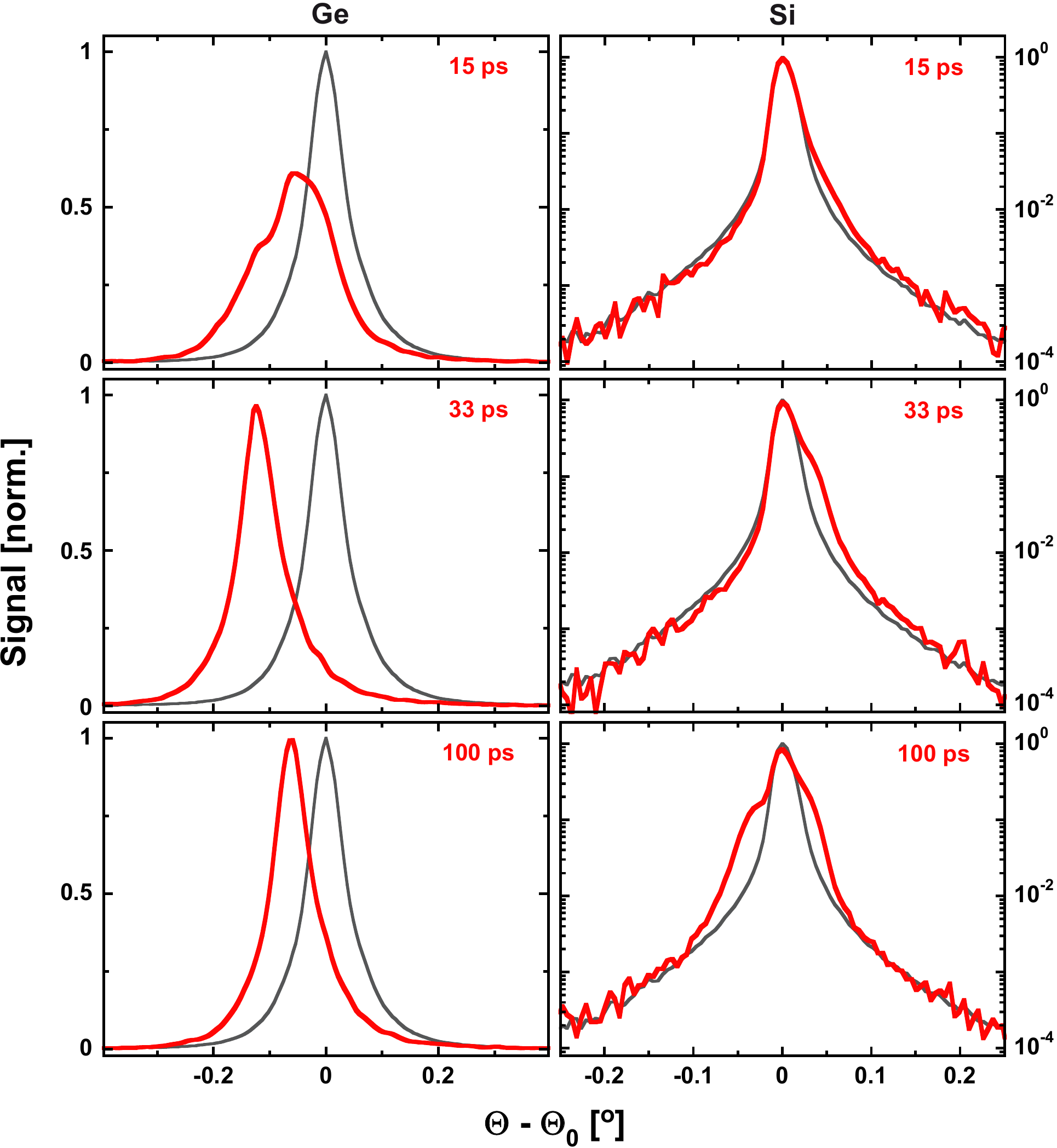}
\caption{Transient rocking curves (red) of the (111) Bragg reflection of (i) a 180 nm Ge film (left, linear scale) and (ii) the (111)-reflection of the bulk Si substrate (right, logarithmic scale) for different pump-probe delay times. Gray curves: Experimental data measured at $\Delta t = -15$ ps as reference. Red curves: experimental data at different delays. \label{fig8}}
\end{figure}

The current configuration, by using the bent crystal optics, allows experiments with high angular/momentum resolution and is - as demonstrated by the data presented in Fig.\ \ref{fig8} - well suited to monitor transient changes of rocking curves. However, the modular approach provides a high flexibility to adopt the set-up to specific requirements of a particular experiment/application: (i) The target material defines the X-ray photon energy and we use our wire-source also with Cu (E$_{K\alpha}$ = 8.05 keV) since suitable X-ray optics (bent crystals\cite{nicoul05}, multilayer optics\cite{bargheer05,shymanovich08}) are available. (ii) Using multi-layer optics, which exhibit a significantly larger bandwidth (i.e.\ full K$_\alpha$ emission) we can expect with our current source an almost an order-of-magnitude higher K$_\alpha$ photon flux on the sample. Such a configuration can be used when the shape and position of the rocking curve does not change, but only the diffraction intensity due to structure factor changes (e.g.\ caused by the excitation of optical phonons\cite{sokolowski03, fritz07}). (iii) Also the sample environment can be flexibly changed to allow for example measurements at low temperatures (cryostat) or the study of irreversible dynamics (e.g.\ melting\cite{siders99, rousse01, sokolowski01}) which require a sample manipulator/goniometer for large samples ($\approx$ 10 cm) and rapid sample motion since a fresh sample area has to be provided for each pulse.

It also needs be stressed that many other laser plasma X-ray sources employ few-mJ, kHz repetition rate laser systems and achieve a similar average X-ray flux. In contrast, we use a high pulse energy (> 100 mJ), low repetition rate drive laser, which results in a two orders of magnitude higher per-pulse X-ray flux. Therefore, a correspondingly lower number of X-ray probe- and optical pump cycles is required to obtain time-resolved diffraction data with a similar integrated signal. This reduction of the optical "dose" is critical for the above mentioned studies of irreversible dynamics since sample area is usually limited and more generally when working in an excitation regime where accumulative sample damage becomes an issue.

\begin{acknowledgments}
Financial support by the {\it Deutsche Forschungsgemeinschaft} (DFG, German Research Foundation) through project C01 {\it Structural Dynamics in Impulsively Excited Nanostructures} of the Collaborative Research Center SFB 1242 {\it Non-Equilibrium Dynamics of Condensed Matter in the Time Domain} (project number 278162697) is gratefully acknowledged. The authors thank M.\ Horn-von Hoegen, M.\ Kammler, and T.\ Wietler for providing the Ge-Si-heterostructure used in this work.
\end{acknowledgments}

\newpage

\newpage


\begin{thebibliography}{86}
\bibitem [{cal()}]{calib}%
  \BibitemOpen
  \href@noop {} {}\bibinfo {note} {The quantum efficiency of the CCD was
  measured with a calibrated Si reference detector (Amptek
  XR100-CR).}\BibitemShut {Stop}%
\bibitem [{cal()}]{yield}%
  \BibitemOpen
  \href@noop {} {}\bibinfo {note} {For this measurement the CCD has been placed between the toroidal mirror and the sample and the total K$_{\alpha}$ yield was estimated by taking into account the efficiency of the bent mirror of about $1.8\times 10^{-5}$ (see section \ref{sec:static}) and that the K$_{\alpha2}$ flux is half of the K$_{\alpha1}$ flux (Fig. \ref{fig2}(b)). }\BibitemShut {Stop}%
\bibitem [{cal()}]{prepulse}%
  \BibitemOpen
  \href@noop {} {}\bibinfo {note} {The measurements in Fig.\ \ref{fig3}(a) and (b) have been performed on different days. Day-to-day variations of the source performance are responsible for the slight difference in yield measured at optimum "pre-" and "main-pulse" delay during the lens position scan (blue data points in Fig.\ \ref{fig3}(a)) and the maximum yield reported in Fig.\ \ref{fig3}(b).}\BibitemShut {Stop}%
\end{thebibliography}
\end{document}